\documentstyle[prl,aps,twocolumn,epsfig]{revtex}

\begin{document}

\twocolumn[\hsize\textwidth\columnwidth\hsize\csname @twocolumnfalse\endcsname

\title{{\bf Origin of the high piezoelectric response in PbZr$_{1-x}$Ti$_{x}$O$_{3}$}}
 
\author{R. Guo$^1$, L.E. Cross$^1$, S-E. Park$^1$, 
B. Noheda\thanks{corresponding author, noheda@bnl.gov}$^{2,3}$, D.E. Cox$^3$,
and G. Shirane$^3$}
\address{$^1$Mat. Res. Lab., Pennsylvania State University, PA 16802-4800 }
\address{$^2$Universidad Autonoma de Madrid, 28049-Madrid, Spain}
\address{$^3$Brookhaven National Laboratory, Upton, NY 11973-5000}
\maketitle

\begin{abstract}
High resolution x-ray powder diffraction measurements on poled PbZr$_{1-x}$Ti%
$_{x}$O$_{3}$ (PZT) ceramic samples close to the rhombohedral-tetragonal
phase boundary (the so-called morphotropic phase boundary, MPB) have shown
that for both rhombohedral and tetragonal compositions, the piezoelectric
elongation of the unit cell does not occur along the polar directions but
along those directions associated with the monoclinic distortion. 
This work provides the first direct evidence for the origin of the very high piezoelectricity
in PZT.
\end{abstract}


\vskip1pc]

\narrowtext
\preprint{cond-mat/000-ms}

\newpage

The ferroelectric PbZr$_{1-x}$Ti$_{x}$O$_{3}$ (PZT) system has been
extensively studied because of its interesting physical properties close to
the morphotropic phase boundary (MPB), the nearly vertical phase boundary
between the tetragonal and rhombohedral regions of the phase diagram close
to x= 0.50, where the material exhibits outstanding electromechanical
properties \cite{Jaffe}. The existence of directional behavior for the
dielectric and piezoelectric response functions in the PZT system has been 
predicted by Du 
{\it et al.} \cite{Xiao-hong Du2},\cite{Xiao-hong Du} from a phenomenological approach 
\cite{Haun}. These authors showed that for rhombohedral compositions the piezoelectric response
should be larger for crystals oriented along the [001] direction than for
those oriented along the [111] direction. Experimental confirmation of this 
prediction was obtained \cite{Kuwata,Park,Durbin} 
for the related ferroelectric relaxor system PbZn$_{1/3}$Nb$_{2/3}$-PbTiO$_{3}$ (PZN-PT), 
which has a rhombohedral-to-tetragonal MPB similar to that of PZT, but it has not been 
possible to verify similar behavior in PZT due to the lack of single crystals. Furthermore,  
{\it ab initio} calculations based on the assumption of tetragonal symmetry, that have been 
successful for calculating the piezoelectric properties of pure PbTiO$_3$ \cite{Rabe,Saghi-Szabo2,Saghi-Szabo,Bellaiche},
 were unable to account for the much larger piezoelectric response in PZT compositions 
close to the MPB. Thus, it is clear that the current theoretical models lack some ingredient 
which is crucial to understanding the striking piezoelectric behavior of PZT.

The stable monoclinic phase recently discovered in the ferroelectric PbZr$%
_{1-x}$Ti$_{x}$O$_{3}$ system (PZT) close to the MPB\cite{Noheda1}, provides
a new perspective to view the rhombohedral-to-tetragonal phase
transformation in PZT and in other systems with similar phase boundaries 
as PMN-PT and PZN-PT \cite{Noheda2}. This phase plays a key role in
explaining the high piezolectric response in PZT and, very likely, in other systems with similar MPBs.
 The polar axis of this
monoclinic phase is contained in the (110) plane along a direction between
that of the tetragonal and rhombohedral polar axes \cite{Noheda1}. An
investigation of several compositions around the MPB has suggested a
modification of the PZT phase diagram \cite{Jaffe} as shown in Fig. 1(top
right) \cite{Noheda2}.

\begin{figure}[h]
\epsfig{width=0.80 \linewidth,figure=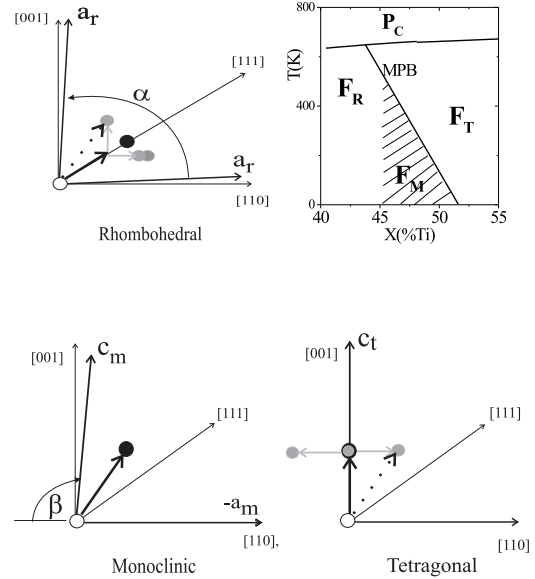}
\vskip2pc
\caption{Schematic view of the PZT phase diagram in the vicinity of the MPB
showing the monoclinic region (top right). A projection of the rhombohedral
(top left), monoclinic (bottom left) and tetragonal (bottom right) unit
cells on the pseudo-cubic (110) plane is sketched. The solid circles
represent the observed lead shifts with respect to the ideal cubic structure
and the grey circles the locally-disordered shifts, four in the tetragonal
phase and three in the rhombohedral phase. The heavy dashed arrows
represent the freezing-out of one of these shifts to give the observed
long-range monoclinic structure \protect\cite{Noheda2}.}
\label{fig1}
\end{figure}

A local order different from the long-range order in the rhombohedral and tetragonal
 phases has been proposed from a
detailed structural data analysis. Based on this, a model has been
constructed in which the monoclinic distortion (Fig. 1, bottom-left) can be
viewed as either a condensation along one of the $\langle $110$\rangle $
directions of the local displacements present in the tetragonal phase \cite{Noheda2} (Fig.
1 bottom-right), or as a condensation of the local displacements along one
of the $\langle $100$\rangle $ directions present in the rhombohedral phase \cite
{Corker} (Fig. 1 top-left). The monoclinic structure, therefore, represents a bridge
between these two phases and provides a microscopic picture of the MPB
region \cite{Noheda2}.

In the present work experimental evidence of an enhanced elongation along
[001] for rhombohedral PZT and along [101] for tetragonal PZT ceramic disks
revealed by high-resolution x-ray diffraction measurements during and after 
the application of an electric field is presented. This experiment was
originally designed to address the question whether poling in the MPB region
would simply change the domain population in the ferroelectric material, or
whether it would induce a permanent change in the unit cell. As shown below,
from measurements of selected peaks in the diffraction patterns, a series of changes
in the peak profiles from the differently oriented grains are revealed which provide
key information about the PZT problem.

PbZr$_{1-x}$Ti$_{x}$O$_{3}$ ceramic samples with x= 0.42, 0.45 and 0.48
were prepared by conventional solid-state reaction techniques using high
purity (better than 99.9\%) lead carbonate, zirconium oxide and titanium
oxide as starting compounds. Powders were calcined at 900$^{o}$C for six
hours and recalcined as appropriate. After milling, sieving, and the
addition of the binder, the pellets were formed by uniaxial cold pressing.
After burnout of the binder, the pellets were sintered at 1250$^{o}$C in a
covered crucible for 2 hours, and furnace-cooled. During sintering, PbZrO$%
_{3}$ was used as a lead source in the crucible to minimize volatilization
of lead. The sintered ceramic samples of about 1 cm diameter were ground to
give parallel plates of 1 mm thickness, and polished with 1 $\mu $m diamond
paste to a smooth surface finish. To eliminate strains caused by grinding
and polishing, samples were annealed in air at 550$^{o}$C for five hours and
then slow-cooled. Silver electrodes were applied to both surfaces of the
annealed ceramic samples and air-dried. Disks of all compositions were poled
under a DC field of 20 kV/cm at 125$^{o}$C for 10 minutes and then
field-cooled to near room temperature. The electrodes were then removed
chemically from the x= 0.42 and 0.48 samples. For the x= 0.45 sample (which had been 
ground to a smaller thickness, about 0.3 mm), the electrodes were retained, so that 
diffraction measurements could be carried out under an electric field.

Several sets of high-resolution synchrotron x-ray powder diffraction
measurements were made at beam line X7A at the Brookhaven National
Synchrotron Light Source. A Ge(111) double-crystal monochromator was used in
combination with a Ge(220) analyser, with a wavelength of about 0.8 \AA ~ in
each case. In this configuration, the instrumental resolution, $\Delta
2\theta $, is an order-of-magnitude better than that of a conventional
laboratory instrument (better than 0.01$^{o}$ in the $2\theta $ region 0-30$%
^{o}$). The poled and unpoled pellets were mounted in symmetric reflection
geometry and scans made over selected peaks in the low-angle region of the
pattern. It should be noted that since lead is strongly absorbing, the
penetration depth below the surface of the pellet at $2\theta =20^{o}$ is
only about 2 $\mu $m. In the case of the x= 0.45 sample, the diffraction
measurements were carried out with an electric field applied in-situ via the
silver electrodes.

\begin{figure}[h]
\epsfig{width=0.80 \linewidth,figure=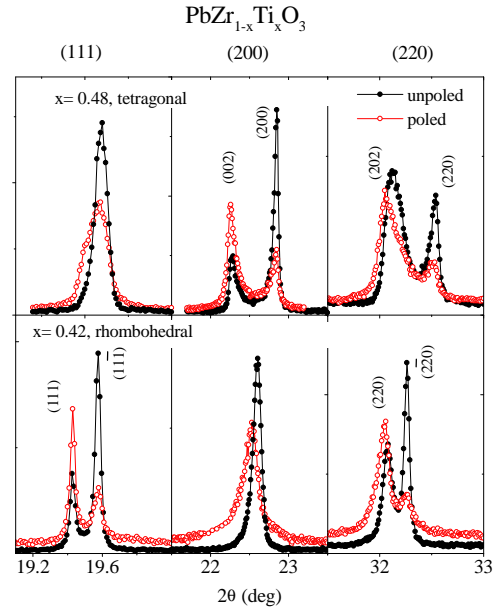}

\caption{(Color) Comparison of (111), (200) and (220) pseudo-cubic reflections for
the x= 0.48 (tetragonal), and x= 0.42 (rhombohedral) PZT compositions before
and after poling}
\label{fig2}
\end{figure}

Powder diffraction measurements on a flat plate in symmetric reflection, in 
which both the incident and the diffracted wave vectors are at the same angle, 
$\theta $, with the sample plate, ensures that the scattering vectors are
perpendicular to the sample surface. Thus only crystallites with their scattering 
vector parallel to the applied electric field are sampled. Scans over selected 
regions of the diffractogram, containing the (111), (200) and (220) pseudo-cubic 
reflections, are plotted in Fig.\ref{fig2} for poled and unpoled PZT samples with the
compositions x= 0.48 (top) and x= 0.42 (bottom), which are in the tetragonal and 
rhombohedral region of the phase diagram, respectively. The diffraction profiles of the poled
and unpoled samples show very distinctive features. For the tetragonal
composition (top), the (200) pseudo-cubic reflection (center) shows a large increase in
the tetragonal (002)/(200) intensity ratio after poling due to the change in the domain 
population, which is also reflected in the increased (202)/(220) intensity ratio in the
 right side of the figure. In the rhombohedral composition with x= 0.42 
(bottom of Fig.\ref{fig2}), the expected change in the domain population can be observed 
from the change of the intensity ratios of the rhombohedral (111) and (11$\overline{1}$) 
reflections (left side ) and the (220) and (2$\overline{2}$0) reflections (right side).

In addition to the intensity changes, the diffraction patterns of the poled
samples show explicit changes in the peak positions with respect to the unpoled 
samples, corresponding to specific alterations in the unit cell dimensions. In the
rhombohedral case (x= 0.42), the electric field produces no shift in the
(111) peak position (see bottom-left plot in Fig.\ref{fig2}), indicating
the absence of any elongation along the polar directions after the application of
the field. In contrast, the poling does produce a notable shift of the
(00l) reflections (center plot), which corresponds to a very significant change of
d-spacing, with $\Delta d/d= 0.32$\%, $\Delta d/d$ being defined as $(d_{p}-d_{u})/d_u$, 
where $d_{p}$ and $d_u$ are the d-spacings of the poled and
unpoled samples, respectively. This provides experimental confirmation
of the behavior predicted by Du {\it et al.} \cite{Xiao-hong Du} for
rhombohedral PZT, as mentioned above. The induced change in the dimensions of
the unit cell is also reflected as a smaller shift in the (202) reflection
(right side plot), corresponding to a $\Delta d/d$ along [101] of 0.12\%.

\begin{figure}[tbp]
\epsfig{width=0.75 \linewidth,figure=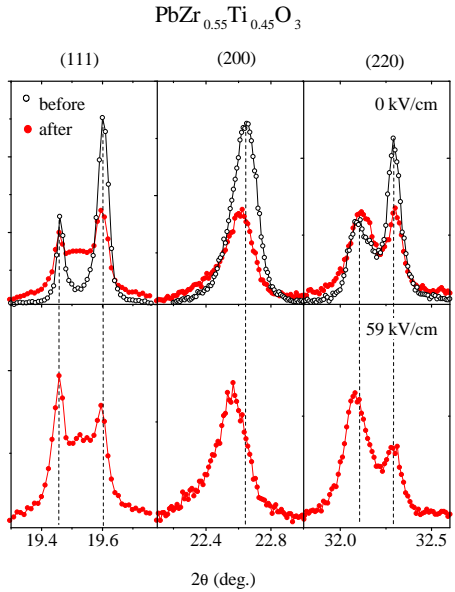}
\caption{(Color), (111), (200) and (220) pseudocubic reflections for PZT with x= 0.45 measured on an unpoled 
sample (open circles) and on a similar sample after the application and removal of a field 
of 59 kV/cm at room temperature (solid circles) are plotted in the upper part of the figure. 
The scattered intensity at $2\protect\theta = 19.52^o$ 
from the second sample corresponds to the (111) reflection from the silver electrode. 
Measurements on the latter sample under an electric field of 59 kV/cm applied {\it in situ} are 
plotted in the lower part of the figure.}
\label{fig3}
\end{figure}

In the tetragonal case for x= 0.48 (top of Fig.\ref{fig2}), there is no peak
shift observed along the polar [00l] direction (center plot), but the 
(202) and the (111) reflections exhibit striking shifts (right and left sides, 
respectively). Furthermore, this composition, which at room temperature is
just at the monoclinic-tetragonal phase boundary, shows, after poling, a
clear tendency towards monoclinic symmetry, in that the (111) and (202)
reflections, already noticeably broadened in the unpoled sample and indicative of an
incipient monoclinicity, are split after poling. These data clearly demonstrate, 
therefore, that whereas the changes induced in the unit cell after the application 
of an electric field do not increase either the rhombohedral or the tetragonal strains, a definite 
elongation is induced along those directions associated with the monoclinic distortion.

In addition to the measurements on the poled and unpoled samples, diffraction measurements 
were performed {\it in situ} on the rhombohedral PZT sample with x= 0.45 as a function of 
applied electric field at room temperature. The results are shown in Fig.3 where the (111), (200) and (220) pseudo-cubic
reflections are plotted with no field applied (top) and with an applied field of 59 kV/cm field (bottom). 
The top part of the figure also shows data taken after removal of the field. As can be seen, 
measurements with the field applied show no shift along the polar [111] direction but, in contrast, 
there is a substantial shift along the [001] direction similar to that for the poled sample 
with x= 0.42 shown in Fig. 2, proving that the unit cell elongation induced by the application 
of a field during the poling process corresponds to the piezoelectric effect induced by the 
in-situ application of a field. Comparison of the two sets of data for x= 0.45 
before and after the application of the field shows that the poling effect of the electric 
field at room temperature is partially retained after the field is removed, although the 
poling is not as pronounced as for the x= 0.42 sample in Fig. 2.

A quantification of the induced microstrain along the different directions
has been made by measuring the peak shifts under fields of 31 and 59 kV/cm.
In Fig. 4, $\Delta d/d$ is plotted {\it versus} the applied
field, $E$, for the (200) and (111) reflections. These data show an approximately linear increase 
in  $\Delta d/d$ for (200) with field, with $\Delta d/d= 0.30\%$ at 59 kV/cm, corresponding 
to a piezoelectric coefficient $d_{33}$$\approx $ 500 pm/V, but essentially no 
change in the d-spacing for (111).

It is interesting to compare in Fig.4 the results of dilatometric measurements of the macroscopic
 linear elongation ($\Delta l/l$) on the same pellet, which must also reflect the effects of
 domain reorientation. At higher fields, this contribution diminishes and one could expect the 
$\Delta l/l$ vs. E curve to fall off between those for the [100] and [111] oriented grains, 
typical of the strain behaviour of polycrystalline ceramics \cite{Park2}. Although such a trend is 
seen above 30 kV/cm, it is intriguing to note that below this value, the macroscopic behavior is
essentially the same as the microscopic behavior for the (200) reflection.

\begin{figure}[tbp]
\epsfig{width=0.90 \linewidth,figure=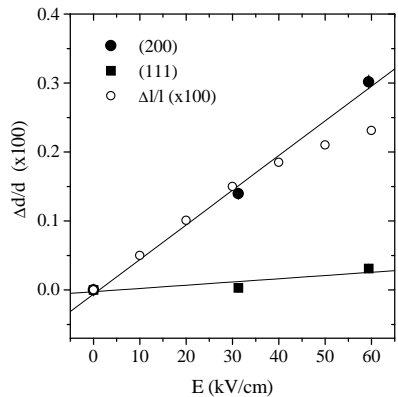}
\caption{Fractional change of d-spacing, $\Delta d/d$ for PZT with x= 0.45 from the rhombohedral (200) 
and (111) reflections as a function of electric field (closed symbols). Dilatometric measurements 
of the macroscopic $\Delta l/l$ for the same pellet are also shown (open circles)}
\label{fig4}
\end{figure}

It is of interest to relate our observations to the more conventional 
description of piezoelectric effects in ceramics, in which the 
dielectric displacements would be attributed to tilts of the polar axis.
What we actually observe in the diffraction experiment is an intrinsic 
monoclinic deformation of the unit cell as a consequence of the rotation of 
the polar axis in the monoclinic plane.
However, large atomic displacements can only occur in 
ompositions close to the MPB, and it is this feature which 
accounts for the sharp peak in the piezoelectric d constants for 
compositions close to 52:48 Zr/Ti \cite{Jaffe}.

We therefore conclude that the piezoelectric strain in PZT close to the morphotropic 
phase boundary, which produces such striking electromechanical properties, is not along 
the polar directions but along those directions 
associated with the monoclinic distortion.  This work supports a model based 
on the existence of local monoclinic shifts superimposed on the rhombohedral and tetragonal
displacements in PZT which has been proposed from a detailed structural analysis of tetragonal \cite{Noheda2} 
and rhombohedral \cite{Corker} PZT samples. Very recent first-principles calculations by 
L. Bellaiche et al. \cite {Bellaiche4} have been
able not only to reproduce the monoclinic phase but also to explain the high piezoelectric coefficients
by taking into account rotations in the monoclinic plane.

As demonstrated above, these high resolution powder data provide key information to
understanding the piezoelectric effect in PZT. In particular, they allow an accurate 
determination of the elongation of the unit cell along the direction of the electric field, 
although they give no information about the dimensional changes occurring along the
perpendicular directions, which would give a more complete characterization of the new
structure induced by the electric field. It is interesting to note that in the case of the related
 ferroelectric system PZN-PT, the availability of single crystals has allowed Durbin {\it et al.} \cite{Durbin} 
to carry out diffraction experiments along similar lines at a laboratory x-ray source. Synchrotron x-ray
experiments by the present authors are currently being undertaken on PZN-PT single
crystals with Ti contents of 4.5 and 8\% under an electric field, and also on other ceramic PZT samples.
Preliminary results on samples with x= 0.46 and
0.47, which are monoclinic at room temperature, have already been obtained. In these cases, the changes 
of the powder profiles induced by poling are so drastic that further work is needed in order to achieve a proper interpretation. \newline

We thank L. Bellaiche, A. M. Glazer, J.A. Gonzalo and K. Uchino for their stimulating
dicussions, B. Jones and E. Alberta for assisting in the sample preparation, 
and A. L. Langhorn for his invaluable technical support. Financial support by
the U.S. Department of Energy under contract No. DE-AC02-98CH10886, and by ONR
under project MURI (N00014-96-1-1173) is also acknowledged.


\begin{references}

\bibitem{Jaffe}  B. Jaffe, W.R. Cook, and H. Jaffe, Piezoelectric Ceramics,
Academic Press, London (1971).
\bibitem{Xiao-hong Du2}  X-h Du, U. Belegundu, and K. Uchino, Jpn. J. Appl. 
Phys. {\bf 36}, 5580 (1997).
\bibitem{Xiao-hong Du}  X-h Du, J. Zheng, U. Belegundu, and K. Uchino, Appl.
Phys. Lett {\bf 72}, 2421 (1998).
\bibitem{Haun} M. J. Haun, E. Furman, S-J. Jang, and L.E. Cross, Ferroelectrics 
{\bf 72}, 13(1989).  
\bibitem{Kuwata}  J. Kuwata, K. Uchino, and S. Nomura, Jpn. J. Appl. Phys. {\bf 21}, 
1298(1982).
\bibitem{Park}  S-E. Park and T. R. Shrout, J. Appl. Phys. {\bf 82}, 1804
(1997).
\bibitem{Durbin}  M. K. Durbin, E.W. Jacobs, J.C. Hicks, and S.-E. Parks,
Appl. Phys. Lett. 74, 2848 (1999).
\bibitem{Rabe} K.M. Rabe and E. Cockayne, in {\it First-Principles Calculations for 
Ferroelectrics: Fifth Williamsburg Workshop}, edited by R.E. Cohen (AIP, Woodbury,1998), 
p. 61.
\bibitem{Saghi-Szabo2}  G. Saghi-Szabo, R.E. Cohen, and H. Krakauer, Phys.
Rev. Lett.  {\bf 80}, 4321 (1998).
\bibitem{Saghi-Szabo}  G. Saghi-Szabo, R.E. Cohen, and H. Krakauer, Phys.
Rev. B {\bf 59}, 12771 (1999).
\bibitem{Bellaiche}  L. Bellaiche and D. Vanderbilt, Phys. Rev. Lett. {\bf 83%
}, 1347 (1999).
\bibitem{Noheda1}  B. Noheda, D.E. Cox, G. Shirane, J.A. Gonzalo, L.E.
Cross, and S-E. Park, Appl. Phys. Lett. {\bf 74}, 2059 (1999).
\bibitem{Noheda2}  B. Noheda, J. A. Gonzalo, L.E. Cross, R. Guo, S-E. Park,
D.E. Cox, and G. Shirane, Phys. Rev. B, (01-April-2000) in press, e-print: cond-mat/9910066 .
\bibitem{Corker}  D.L. Corker, A.M. Glazer, R.W. Whatmore, A. Stallard, and
F. Fauth, J. Phys.:Condens. Matter {\bf 10}, 6251 (1998).
\bibitem{Park2} S.-E. Park, T.R. Shrout, P. Bridenbaugh, J. Rottenberg, and G. Loiacono, 
Ferroelectrics {\bf 207}, 519 (1998).
\bibitem{Bellaiche4} L. Bellaiche, A. Garcia, and D. Vanderbilt. To be published.
\end{references}
\end{document}